%% file: main.tex
\documentclass[lettersize,journal]{IEEEtran}
\usepackage[dvipsnames]{xcolor}
\usepackage{amsmath,amsfonts}
\usepackage{algorithmic}
\usepackage{algorithm}
\usepackage{array}
\usepackage[caption=false]{subfig}
\usepackage{textcomp}
\usepackage{stfloats}
\usepackage{url}
\usepackage{verbatim}
\usepackage{graphicx}
\usepackage{cite}
\graphicspath{{figs/}}

\newcommand{\bhl}[1]{{{#1}}}

\begin{document}
\bstctlcite{IEEEexample:BSTcontrol}

\title{A 20-Year Retrospective on Power and Thermal Modeling and Management}


\author{%
\IEEEauthorblockN{David~Atienza,~\IEEEmembership{Fellow,~IEEE}\IEEEauthorrefmark{1}, Kai~Zhu\IEEEauthorrefmark{1}, Darong~Huang\IEEEauthorrefmark{1}, Luis~Costero\IEEEauthorrefmark{2}}
\\\IEEEauthorblockA{\IEEEauthorrefmark{1}%
\{david.atienza, kai.zhu, darong.huang\}@epfl.ch\\
Embedded Systems Laboratory, EPFL, Switzerland}
\\\IEEEauthorblockA{\IEEEauthorrefmark{2}lcostero@ucm.es\\
Universidad Complutense de Madrid, UCM, Spain }
}

\markboth{IEEE Design \& Test}%
{Shell \MakeLowercase{\textit{et al.}}: A Sample Article Using IEEEtran.cls for IEEE Journals}


\maketitle

\begin{abstract}
As processor performance advances, increasing power densities and complex thermal behaviors threaten both energy efficiency and system reliability. This survey covers more than two decades of research on power and thermal modeling and management in modern processors. We start by comparing analytical, regression-based, and neural network-based techniques for power estimation, then review thermal modeling methods, including finite element, finite difference, and data-driven approaches. Next, we categorize dynamic runtime management strategies that balance performance, power consumption, and reliability. Finally, we conclude with a discussion of emerging challenges and promising research directions.
\end{abstract}


\input{s1_introduction}

\input{s2_power_modeling}
\input{s3_thermal_modeling}
\input{s4_power_thermal_management}

\input{s5_future_challenges}


\section*{Acknowledgment}
This work was supported in part by the Swiss State Secretariat for Education, Research, and Innovation (SERI) through the SwissChips research project, and by Intel as part of the Intel Center for Heterogeneous Integrated Platforms (HIP).
This work was supported in part by Grant PID2021-126576NB-I00 funded by MCIN/AEI/10.13039/501100011033 and by ``ERDF A way of making Europe''

\bibliographystyle{IEEEtran}

\input{bio}

\vfill

\end{document}

%% file: s1_introduction.tex
\section{Introduction}

During the past few decades, microprocessor performance has rapidly advanced following Moore's law, driven by shrinking device dimensions and increasing transistor densities \cite{hennessy2017computer}. This progress was long sustained by Dennard scaling, which kept power density roughly constant as transistors became smaller. However, the breakdown of Dennard scaling around 2005 led to the emergence of a `power wall' \cite{hennessy2017computer}, limiting further increases in clock frequency. To sustain performance growth under these power constraints, chip designers shifted towards multi-core architectures, distributing workloads across multiple lower frequency cores instead of solely increasing single core frequency \cite{li2009mcpat}. However, multi-core scaling also faces limits: as core counts rise, cumulative power consumption and heat generation prevent all cores from running at full frequency simultaneously. This leads to the `dark silicon' phenomenon, where some cores must remain inactive or run at lower frequencies to stay within power and thermal budgets \cite{huang2024evaluation}. More recently, heterogeneous architectures have emerged that integrate general-purpose cores with specialized accelerators such as graphics processing units (GPUs) to offer better power efficiency (performance per watt) \cite{li2009mcpat,zhou2019primal}. Given these ongoing power challenges, precise power modeling and dynamic management strategies are crucial for monitoring and controlling on-chip power dissipation. 



As transistor scaling approaches physical limits, following Moore's law has become increasingly challenging. Advanced packaging techniques, such as 2.5D chiplet systems and 3D integrated circuits (ICs) \cite{ladenheim2018mta}, have been developed to further enhance integration. In a 2.5D chiplet design, multiple chiplets are positioned side-by-side on an interposer, providing much finer wiring pitches and higher die-to-die bandwidth. 3D ICs vertically stack dies using through-silicon vias (TSVs) to enable shorter interconnect lengths and improve performance \cite{ladenheim2018mta}. Although these techniques improve integration, they concentrate power dissipation in smaller volumes and create complex thermal pathways \cite{terraneo20213d}. These factors present significant challenges for thermal modeling and management to maintain system performance and reliability \cite{walker2016accurate,huang2022reinforcement}.

Throughout the semiconductor development process, power and thermal considerations are as crucial as performance, making precise modeling indispensable both at design-time and during runtime \cite{terraneo20213d, huang2024evaluation}. In the design phase, power and thermal models assist designers in designing space exploration (DSE) and ensure that candidate designs meet the requisite performance and power budgets before fabrication \cite{li2009mcpat}. At runtime, chips exhibit dynamic power and thermal behaviors depending on workload and environmental conditions, requiring runtime power and thermal estimators to monitor application execution. Then, effective policies are needed to adjust operating parameters to improve energy efficiency and reliability. 

\begin{figure}[h]
  \centering
  \includegraphics[width=0.45\textwidth]{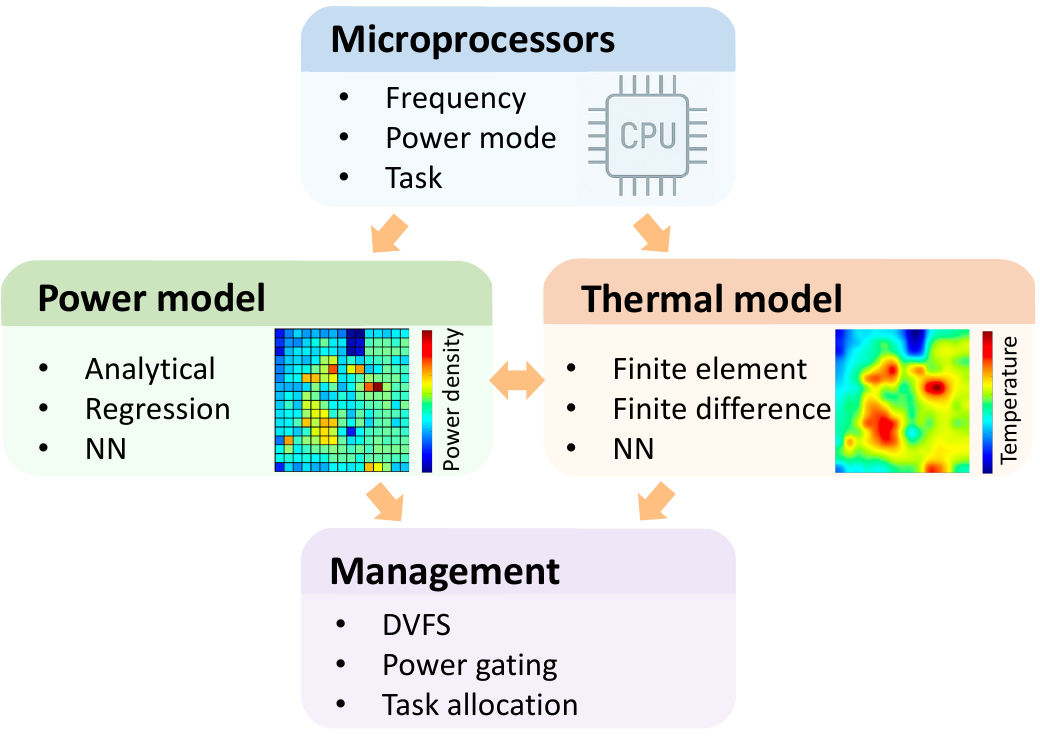}
  \caption{Workflow of power modeling, thermal modeling, and management.}
  \label{outline_workflow}
\end{figure}


It is essential to understand the interactions between power and thermal modeling and management, which is illustrated in Fig. \ref{outline_workflow}. Power models provide critical input for thermal analysis, as nearly all electrical power eventually dissipates as heat. In contrast, temperature variations influence power consumption, notably through temperature-dependent leakage power. Power and thermal profiles form the quantitative foundation for runtime management policies. Based on this workflow, this review is structured around four interrelated research areas:
\begin{itemize}
  \item Power modeling techniques for processors, including analytical models, regression-based approaches, and neural network (NN) models.
  \item Thermal modeling methods for capturing on-chip temperature responses, encompassing numerical methods and neural network techniques.
  \item Dynamic power and thermal management strategies to balance performance, power consumption, and on-die temperature distribution.
  \item Emerging trends and future challenges, spanning power models, advanced cooling solutions, detailed thermal simulators, holistic multi-physics co-simulation frameworks, and runtime control for heterogeneous multi-processor system-on-chips (MPSoCs).
\end{itemize}

%% file: s2_power_modeling.tex
\section{Power modeling}

Processor power consumption can be measured directly using on-die power sensors or external instruments. However, on-die power sensors suffer from three primary limitations \cite{walker2016accurate}: (1) restricted spatial and temporal resolution; (2) lack of flexibility, since the number and placement of sensors are usually fixed at design time; (3) scalability challenges due to physical size and deployment cost. External power meters can capture the total power draw of the whole system, but detailed power breakdowns for individual components cannot be obtained. To overcome these limitations, researchers have developed a range of power-modeling techniques.

\subsection{Analytical models}
Analytical power models for CMOS circuits break down total power dissipation into dynamic and static/leakage components. Dynamic power includes switching power and short-circuit power. The predominant switching power relates to the energy required to charge and discharge the load capacitances and can be expressed as
\begin{equation}
P_{\text {dynamic }}= \alpha \cdot f  \cdot  C \cdot V^2 
\label{eq:dynamic_power}
\end{equation}
where $\alpha$ is the activity factor, $f$ is the switching frequency, $C$ is the load capacitance, and $V$ denotes the supply voltage.
Static power arises from leakage currents that persist even when transistors are off, and is given by
\begin{equation}
P_{\text {static}}= I_{\text{leakage}} \cdot V 
\label{eq:static_power}
\end{equation}
where $I_{\text{leakage}}$ aggregates sub-threshold, gate-oxide and junction leakage currents.

Analytical power models can be categorized by their abstraction levels, from gate level and register-transfer level (RTL) up to architecture level. As the modeling abstraction increases, the simulation speed increases at the expense of estimation fidelity. 

\begin{figure}[htb]
  \centering
  \subfloat[Gate level]{%
    \includegraphics[width=0.33\linewidth]{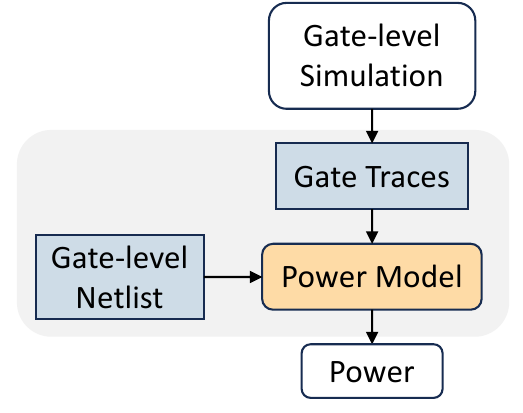}%
  }\hfill
  \subfloat[RTL level]{%
    \includegraphics[width=0.33\linewidth]{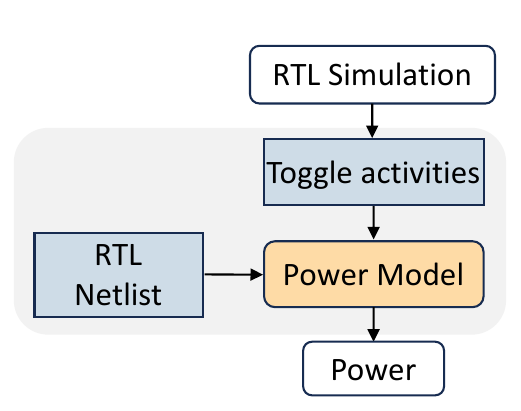}%
  }\hfill
  \subfloat[Architecture level]{%
    \includegraphics[width=0.33\linewidth]{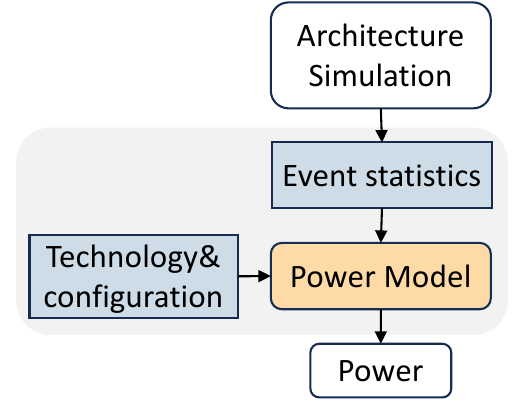}%
  }
  \caption{Analytical models of different abstraction levels.}
  \label{power_abstract_level}
\end{figure}

\subsubsection{Gate level}

\bhl{Gate-level power analysis tools like} Synopsys PrimeTime PX (PTPX) take a gate-level netlist along with switching activity data from simulation, as shown in Fig. \ref{power_abstract_level} (a). The tool maps each net’s capacitance and toggle rate to compute $P_{\text {dynamic}}$ and sums each cell's leakage currents to estimate $P_{\text {static}}$. PTPX can provide cycle-accurate power traces and is often considered a gold reference for power analysis. However, exhaustive gate-level analysis is time-consuming, often taking hours or even days to evaluate a single intellectual property (IP) core \cite{zhou2019primal}. 




\subsubsection{RTL level}
At the register-transfer level, tools such as Ansys PowerArtist and Siemens PowerPro enable faster early-stage power estimation and optimization. As shown in Fig. \ref{power_abstract_level} (b), they use the toggle activity captured during RTL simulation and combine it with ready‑made cell‑and‑wire models from the technology library to estimate both dynamic and static power. Although RTL power analysis runs faster than gate-level analysis, it still struggles with large designs or long simulation traces. This motivates researchers to develop architecture-level power models that prioritize speed and enable real-time power monitoring. 



\subsubsection{Architecture level}
Architecture-level models abstract low-level circuit behaviors into high-level hardware events and parameters for rapid power estimation, as shown in Fig. \ref{power_abstract_level} (c). A classical example is Wattch, proposed by Brooks et al. \cite{brooks2000wattch}, which estimates the dynamic power of functional processor units. Wattch computes the load capacitance of each unit, and derives activity factors from architectural simulators such as Gem5, then aggregates the dynamic power across all units. To improve accuracy, subsequent tools, such as CACTI, incorporate leakage power estimation. For multi-core processors, simply summing the power of individual cores can be inaccurate due to shared resources and inter-core contention \cite{li2009mcpat}. The McPAT framework \cite{li2009mcpat} addresses this by modeling shared components, contention effects, and other overheads, yielding more accurate power estimates for multi-core architectures. 

Despite these advancements, studies have found that McPAT's power estimates can deviate by up to 40\% from measured values, primarily due to modeling abstractions. To mitigate this error, researchers have applied data-driven calibration techniques, such as McPAT-Calib~\cite{zhai2022mcpat}, which calibrates McPAT against real measurements. However, such calibrated models are tuned to specific microarchitectures, which limits their generalizability across different processor designs \cite{zhai2023microarchitecture}. 


\subsection{Regression models}

In addition to analytical models, regression models based on architecture-level activity metrics have been developed for runtime power estimation \cite{bertran2013counter}. These models typically express power consumption as a weighted sum of event counts collected from on-chip performance monitor counters (PMCs).

PMC-based regression models provide notable advantages due to their low computational overhead and ease of implementation. However, the underlying relationship between raw event counts and power can be complex and non-linear. As a result, recent work has increasingly turned to neural network models that can capture intricate interactions between system activity and power consumption.

\subsection{Neural network-based models}

Early works used relatively simple multilayer perceptrons (MLPs) trained on performance counter data. For instance, Walker et al. \cite{walker2016accurate} developed a shallow MLP model that achieves millisecond-scale inference latency, and the accuracy outperforms linear regression models under non-stationary workloads by capturing non-linear relationships. More recent efforts have expanded simple MLP systems into more complicated network architectures to improve fidelity. Zhou et al. \cite{zhou2019primal} proposed PRIMAL, a convolutional neural network (CNN) model trained on flip-flop signal data to predict per-cycle power consumption. Although PRIMAL achieves high temporal resolution and accuracy, it requires extensive training data and significantly longer training time than regression methods. Zhang et al. \cite{zhang2020grannite} introduced GRANNITE, which leverages a graph neural network (GNN) to represent gate-level netlists as graphs and applies message-passing layers to predict average power for given workloads. \bhl{These approaches demonstrate the potential of neural networks to model power, especially for capturing complex patterns that regression models can miss.}

%% file: s3_thermal_modeling.tex
\section{Thermal modeling}

Since nearly all power consumed by a processor is ultimately converted into heat, the power maps derived in the previous section serve as the heat source inputs for thermal analysis. Accurate on-chip temperature profiling is essential for evaluating system reliability and performance. On-die sensors can provide direct temperature measurements, but their spatial resolution is limited due to area and cost constraints \cite{li2020accurate}. IR thermography can capture high-resolution temperature maps of the surface of a chip, but it can be obstructed by packaging components such as heat spreaders or heat sinks and may not accurately reflect internal junction temperatures \cite{sadiqbatcha2020post}. Given these limitations of direct measurement, numerical methods are widely used to predict on-die temperature distributions.





\subsection{Heat transfer equation}

Numerical methods estimate the temperature distribution by solving the governing heat transfer equation. In microprocessors, heat spreads primarily via conduction through solid materials and convection at interfaces between solids and surrounding fluids, such as air or liquid coolants. The governing equation for heat conduction \cite{terraneo20213d} is given by
\begin{equation}
\nabla  \cdot (k\nabla T) + \dot{q} = \rho C_v \frac{{\partial T}}{{\partial t}}
\label{eq:heat_transient}
\end{equation}
where $\rho$ is the material density, $C_v$ denotes the specific heat capacity, $k$ represents the thermal conductivity, and $\dot{q}$ is the volumetric heat source. This partial differential equation can be solved with appropriate boundary conditions, such as adiabatic or convective boundary conditions that model heat insulation or heat exchange with the ambient environment. 




\subsection{Numerical methods}

\subsubsection{Finite element method (FEM)}
FEM discretizes the chip's geometry into an unstructured mesh of small elements and approximates the temperature field within each element using local basis functions. The elemental equations are assembled into a global matrix equation, which is then solved after applying boundary conditions to obtain the temperature distribution across the entire domain. FEM can achieve high accuracy for complex geometries, heterogeneous materials, and anisotropic thermal conductivities, and it is widely used in commercial tools like ANSYS and COMSOL. However, FEM is computationally expensive, \bhl{and not suitable for run-time thermal management \cite {pfromm2024mfit}}. 
To improve efficiency, researchers have proposed several enhancements. For example, in \cite{ladenheim2018mta} an adaptive meshing strategy is employed to refine the mesh locally in regions with steep temperature gradients, reducing the number of elements without sacrificing precision. In \cite{lu2016electrical}, domain decomposition methods (DDM) are employed to partition the domain into subdomains that can be meshed independently and solved in parallel. This approach not only simplifies the generation of meshes for complex layouts but also enhances both computational flexibility and efficiency. 



\subsubsection{Finite difference method (FDM)}
\begin{figure}[h]
  \centering
  \includegraphics[width=0.45\textwidth]{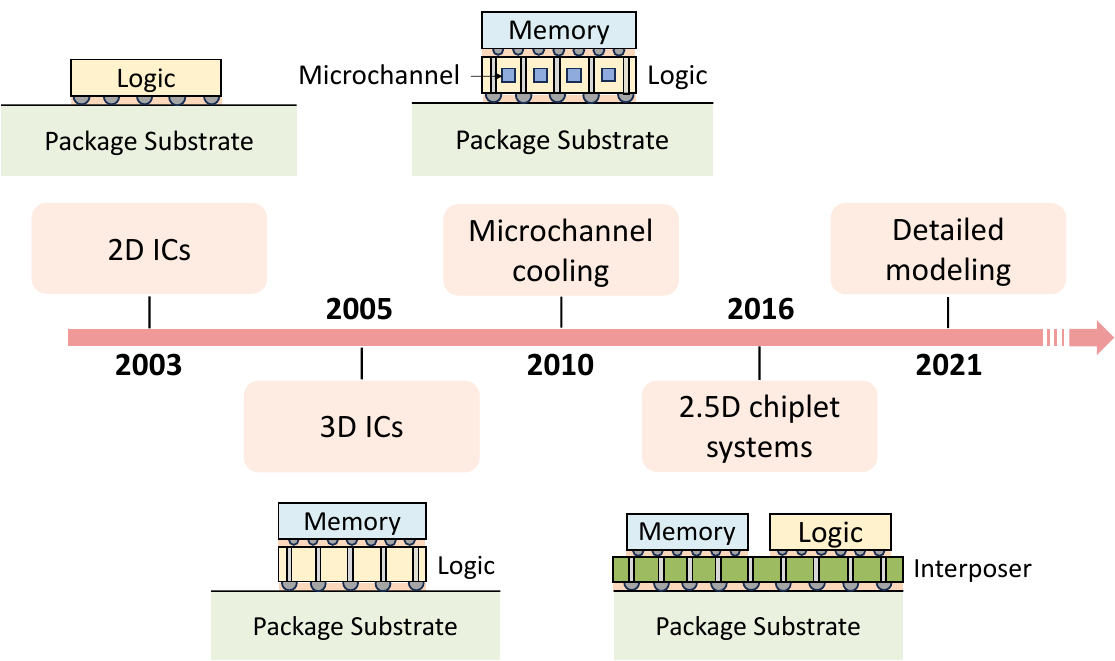}
  \caption{Development of compact thermal models.}
  \label{thermal_time}
\end{figure}

FDM discretizes the chip into structured grids, which is simpler to implement and computationally cheaper than FEM, \bhl{therefore widely used for run-time thermal management \cite{pfromm2024mfit, huang2024evaluation}}. The drawback is that structured grids cannot easily conform to irregular geometries, which reduces the precision for complex layouts. A prominent subclass of FDM is compact thermal models (CTMs), which simplify the heat equation into an equivalent lumped RC network based on the analogy between thermal and electrical properties. 
\bhl{The evolution of CTMs over time is illustrated in Fig. \ref{thermal_time}}. Hotspot~\cite{zhang2015hotspot} was first proposed \bhl{in 2003} for thermal modeling of 2D ICs. It subdivides the chip into uniform grid cells and builds a thermal RC network based on the floorplan and power distribution of each component. The RC network is then solved to obtain the temperature distribution. Later versions of Hotspot~\cite{zhang2015hotspot} added support for multi-layer stacks \bhl{in 2005} and the heterogeneous material properties within individual layers \bhl{in 2016}. 3D-ICE~\cite{terraneo20213d} extends CTMs to 3D ICs with microchannel liquid cooling \bhl{in 2010} by adding voltage-controlled current sources to model convective heat removal by microchannel coolant flows. 3D-ICE also provides an extensible plug-in framework to model other advanced cooling technologies. Since version 3.1, 3D-ICE supports non-uniform grids, allowing the use of finer grids in regions with high temperature gradients and coarser cells elsewhere to balance accuracy and efficiency \cite{huang2024evaluation}. \bhl{PACT~\cite{yuan2021pact} was developed in 2021, which integrates parallel solvers to deliver fast thermal analysis with fine grids required to capture detailed temperature distribution.}




\subsection{Neural network-based methods}

\bhl{Apart from numerical methods}, neural networks have also been explored to reconstruct high-resolution temperature maps from limited data from on-die sensors or PMCs. For example, Li et al. \cite{li2020accurate} train a CNN to exploit spatial correlations in steady-state heat dissipation, successfully inferring full-chip temperature distributions from a dozen sensor readings. Sadiqbatcha et al. \cite{sadiqbatcha2020post} extend this idea to temporal prediction by training a long short-term memory (LSTM) network based on time series of PMC data to predict the transient thermal behavior of multi-core processors. 

%% file: s4_power_thermal_management.tex
\section{Power and thermal management}

Increasing a processor clock frequency can improve performance, but typically requires a higher supply voltage to maintain signal timing margins \cite{hennessy2017computer}. However, a higher voltage results in a disproportionate increase in power consumption and heat generation relative to the performance gain. As indicated in equation~\eqref{eq:dynamic_power}, dynamic power consumption scales roughly with the cube of the supply voltage since frequency scales approximately linearly with voltage within the typical operating range. Therefore, effective management strategies are essential to dynamically adjust processor operating states to achieve high performance while mitigating power and thermal issues~\cite{huang2022reinforcement,huang2024evaluation}.

\subsection{Control methods}

\subsubsection{Dynamic Voltage and Frequency Scaling (DVFS)}

DVFS is a widely used technique to adjust a processor's supply voltage and operating frequency based on workload demands. By lowering voltage and frequency during periods of low activity, DVFS can effectively reduce dynamic power consumption.



\subsubsection{Clock and power gating}
Clock and power gating reduce power consumption by disabling inactive processor components. Clock gating disables the clock signal to idle functional units, eliminating unnecessary switching activities and saving dynamic power. Power gating goes further by cutting off the power supply to inactive circuitry, which saves both dynamic and leakage power. However, power gating can introduce latency when transitioning back to active states and requires careful control. 


\subsubsection{Task allocation}
The operating system or runtime schedulers can perform power- and thermal-aware task scheduling across processor cores \cite{wang2015efficient}. By intelligently deciding when and where each thread runs, the scheduler complements hardware DVFS and gating techniques for power and thermal management. 


\begin{figure}[h]
  \centering
  \includegraphics[width=0.4\textwidth]{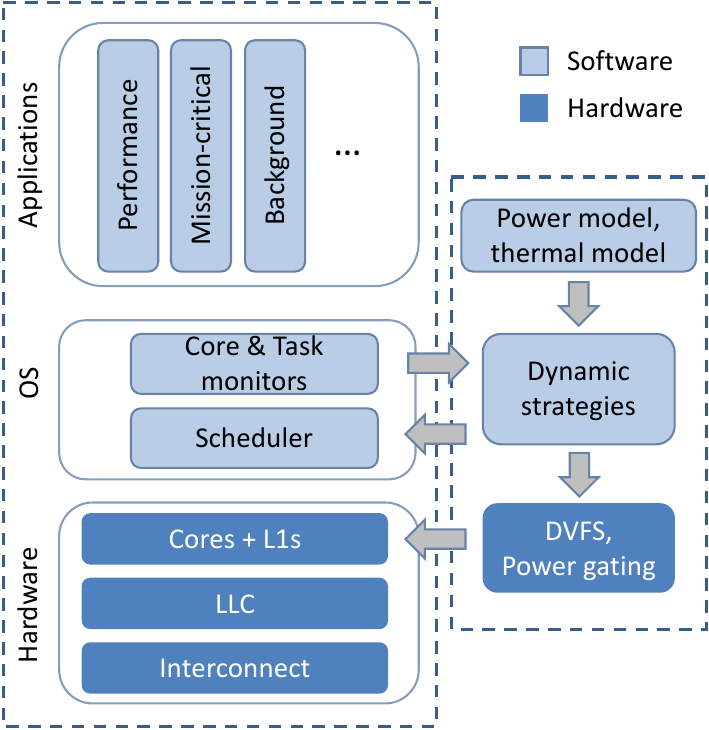}
  \caption{Workflow of power and thermal management.}
  \label{management_workflow}
\end{figure}

\subsection{Dynamic management scenarios}
Dynamic power and thermal management frameworks read data from power and thermal models, core and task monitors, then adapt the processor operation to keep performance, power, and temperature within the target limits, as shown in Fig. \ref{management_workflow}. Approaches in this domain have evolved from heuristic and reactive policies to adaptive and proactive controllers for complex runtime conditions. Early works often rely on stochastic models, linear programming, or rule-based heuristics. More advanced strategies introduce proactive control mechanisms that predict future temperature or power states and adjust settings preemptively~\cite{iranfar2015heuristic}. Recently, reinforcement learning has been integrated into management frameworks, enabling systems to learn optimal policies through adaptive environmental interactions \cite{huang2022reinforcement}. This section presents several runtime management strategies, categorized by their optimization objectives.

\subsubsection{Power minimization under performance \bhl{or real-time} constraints}

Benini et al. \cite{benini1999policy} proposed a reactive power management approach based on a Markov decision process. Their policy selectively transitions processor components between active and idle states to minimize power consumption while satisfying performance requirements.
\bhl{
For real-time systems, Huang et al. \cite{huang2022reinforcement} used Q-learning to dynamically choose each core’s DVFS level and switch the last-level cache (LLC) between fast SRAM and low-leakage RRAM modes. Combined with a preemptive priority queue, their policy cuts both core and cache energy usage while meeting every deadline.
}



\subsubsection{Performance maximization under power or temperature constraints}
Iranfar et al. \cite{iranfar2015heuristic} combined Q-learning with a reduction in the set of heuristic actions for power and thermal management of heterogeneous MPSoCs. This proposed method allows the system to autonomously select DVFS levels and core allocations that increase performance as much as possible while keeping power and temperature below specified limits.



\subsubsection{Peak temperature minimization under performance constraints}
Wang et al. \cite{wang2015efficient} utilized Wattch as the power model and a compact thermal model to predict the temperature distribution in multi-core processors. Their approach manages thermal challenges by employing DVFS to balance task execution times and reduce overall power consumption. Additionally, they introduced a Task Partitioning and Scheduling (TPS) algorithm that divides tasks into hot and cool subtasks, which are interleaved both temporally and spatially across cores to minimize peak temperature without missing performance targets.

%% file: s5_future_challenges.tex
\section{Conclusion and future trends}

\bhl{
Over the past 20 years, power and thermal models for MPSoCs have evolved from analytical and numerical estimates to neural network-based predictors. Meanwhile, runtime management policies have progressed from heuristic controls to learning-driven policies. Looking ahead, power and thermal constraints will continue to be critical bottlenecks for future MPSoCs.}
Keeping improvements in performance and energy efficiency while ensuring system reliability requires further progress in multiple key research directions. These directions include, but are not limited to: (1) Developing accurate power models tailored for diverse components and unified frameworks for power-thermal co-modeling; (2) Exploring advanced cooling strategies to efficiently manage increasing heat densities; (3) Building detailed thermal models that precisely capture complex heat transfer mechanisms; (4) Establishing robust multi-physics co-simulation frameworks capable of integrating electrical, thermal, and mechanical interactions; (5) Enhancing runtime management frameworks that effectively coordinate power and thermal controls across heterogeneous components. 

\subsection{Trends in power modeling}
\subsubsection{Power models for 2.5D/3D heterogeneous MPSoCs}
\bhl{
For 2.5D interposer-based and 3D MPSoCs, heterogeneous tiles like CPUs, GPUs, AI accelerators, high-bandwidth memory (HBM) stacks, and other chiplets are co-integrated within a single package. This diversity requires power models that can capture not only the internal switching behaviors of each component but also interdie phenomena, such as power delivery network interactions, voltage drop effects, and off-die communication overheads. 
Recent advances in deep learning approaches have shown promise in modeling the complex non-linear interactions and cross-die coupling effects that traditional analytical models struggle with \cite{zhou2019primal, zhang2020grannite}. However, these neural network models typically require large and architecture-specific training datasets and often struggle to generalize to new chip designs. To alleviate this data bottleneck, researchers have explored transfer learning techniques. For example, Zhai et al.~\cite{zhai2023microarchitecture} pre-train a neural network on data from a broad range of chip architectures, then fine-tune it using minimal data from a target chip, thus reducing the training effort required for new designs. 
}

Another promising direction is the concept of hybrid modeling, which combines physics-based equations with neural network models. By embedding analytical power formulas into neural networks, hybrid models can reduce the amount of required training data and produce more robust power estimates under diverse operating conditions~\cite{chen2023fast}. In addition, these models tend to be more interpretable and less prone to overfitting, while still capturing complex interactions in heterogeneous systems.

\subsubsection{Power thermal co-modeling}
Beyond standalone power models, power-thermal co-modeling has gained attention due to the tightly coupled feedback loop between power dissipation and temperature, as shown in Fig.~\ref{outline_workflow}. For example, per-core power estimates from \mbox{McPAT} can be fed into 3D-ICE to generate temperature maps. Likewise, PACT~\cite{yuan2021pact} injects gate-level power maps into its thermal model to predict detailed temperature profiles.

Accurate co-modeling requires accounting for temperature-dependent leakage variations. Traditional workflows implement an iterative loop: Start with an initial power map to estimate temperatures, then update the power estimates based on those temperatures, and repeat the cycle until convergence. This iterative loop can be computationally expensive. Therefore, an important research direction is to develop advanced approaches that bypass costly iteration and improve the speed and accuracy of power-thermal co-modeling. 



\subsection{Trends in cooling techniques}
The integration of multiple dies in 3D ICs dramatically increases the power density and vertical thermal resistance, making heat removal much more challenging. Inner dies are susceptible to heat accumulation because of limited thermal pathways, leading to hotspots deep within the package. Conventional air cooling (heat sinks and fans) primarily removes heat from the top of the chip stack, which is insufficient for these embedded hotspots. This has spurred the exploration of advanced cooling techniques such as vapor-chambers, microchannel liquid cooling, immersion cooling, spray- and jet-impingement cooling, and solid-state coolers.

Vapor chambers are flat heat pipes composed of thin copper plates lined with a wick structure and a small amount of working fluid. When a hot chiplet causes the fluid at that location to evaporate, the vapor spreads laterally and condenses at cooler regions, the wick then returns the liquid to hot regions. This mechanism can reduce temperature non-uniformities in heterogeneous MPSoCs. The main challenges in the deployment of vapor chambers include maintaining the proper fluid fill ratio and reliable wick performance over large substrates.

Microchannel cooling embeds tiny fluid channels within the chip to cool hotspots at lower layers. Van Erp et al.~\cite{van2020co} demonstrate that microchannel cooling can dissipate heat fluxes exceeding 1 $\text{kW}/\text{cm}^{2}$ while using very little pumping power, making it one of the most promising near-term enhancements over heat sinks for high-power density devices.
\bhl{At the package level, immersion cooling submerges the entire module in a bath of electrically insulating liquid that removes heat from every exposed surface rather than through etched channels \cite{birbarah2020water}. Two-phase immersion baths have been shown to dissipate local heat fluxes exceeding 500 $\text{W}/\text{cm}^{2}$ in \cite{birbarah2020water}. However, integration hurdles focus on ensuring the compatibility of the coolant with electronic components and designing connectors and seals that function reliably while submerged.}
Spray and jet-impingement cooling techniques focus liquid jets on specific hotspot areas. They can achieve local convective heat transfer coefficients an order of magnitude higher than plain microchannels, and can be retrofitted to existing packages. However, challenges such as nozzle clogging and non-uniform flow distribution pose reliability concerns and complicate qualification processes.

Solid-state cooling devices, such as thermoelectric coolers (TECs) and electrocaloric coolers offer another path to managing on-chip temperatures. These devices can be integrated near hotspots and can actively pump heat without moving fluids. They also have fast response times and small footprints. Recent progress in electrocaloric materials could make efficient solid-state coolers practical, but high costs, integration difficulties, and limited achievable temperature drops confine them currently to laboratory prototypes \cite{liu2024thermoelectric}. 

Future high-performance systems are likely to employ combinations of these advanced cooling methods to handle increasing thermal loads. However, this approach can complicate thermal modeling: microchannels and spray jets add flow-dependent and time-varying convection; vapor chambers introduce orientation-sensitive, anisotropic spreading; and solid-state coolers couple electrical control with thermal effects. Accurate prediction of MPSoC temperatures in such scenarios will demand more sophisticated and flexible thermal models that can account for these dynamic and multidimensional heat transfer phenomena.

\subsection{Trends in thermal modeling}
\subsubsection{Modeling for advanced cooling}
Accurately modeling chips that employ advanced cooling techniques requires versatile and extensible simulation platforms. OpenModelica, an open-source implementation of the Modelica language, is a good candidate for this task and has been integrated into 3D-ICE to model various cooling configurations \cite{terraneo20213d}. \bhl{By using Modelica libraries, researchers can plug in parameterized models for elements such as microchannel flows, pumps with variable speed, or two-phase evaporative cooling sections.} This component-based approach allows complex cooling systems to be constructed and simulated without manually solving the governing equations. Once validated, these cooling component models can be exported as Functional Mock-up Units (FMUs) and co-simulated with thermal simulators such as 3D-ICE. Such co-simulation enables accurate and efficient thermal analysis that accommodates the dynamic behaviors of advanced cooling techniques.


\subsubsection{Detailed compact thermal models}
The anisotropic and intricate heat flow in modern 2.5D/3D ICs challenges conventional coarse-grid models, which might miss localized hotspots or steep temperature gradients, and lead to suboptimal thermal management decisions. To address this, researchers are developing detailed compact thermal models to capture fine-grained heat transfer effects. 

\begin{figure}[h]
  \centering
  \hspace*{-0.5cm}
  \includegraphics[width=0.42\textwidth]{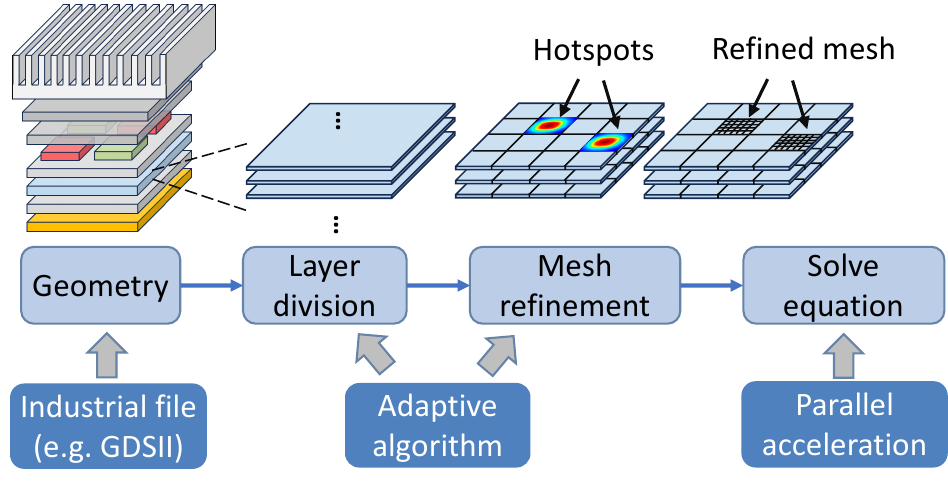}
  \caption{Workflow for detailed thermal modeling in 3D-ICE.}
  \label{3DICE_workflow}
\end{figure}

Fig. \ref{3DICE_workflow} shows this thermal simulation workflow implemented in the latest 3D-ICE. Conventional compact thermal models manually construct chip stacks with each functional layer assumed to be homogeneous. This simplicity delivers fast simulation, while neglecting the heterogeneous and anisotropic characteristics, potentially compromising accuracy. To address this problem, 3D-ICE introduces an interface capable of parsing material and layout details from industrial design files (e.g., GDSII layouts), to automatically build detailed heterogeneous and anisotropic models. This automation reduces manual effort and potential errors, and improves accuracy by accounting for fine-grained material variations. 

A critical aspect of improving compact models is how the chip stack is discretized into layers for the RC network. Traditionally, each functional layer of the chip is modeled as one layer in the network. However, this simplistic division can be inadequate to accurately capture vertical temperature gradients. For example, dielectrics or redistribution layers can exhibit thermal resistance higher than other functional layers, leading to steep vertical temperature gradients that require a finer vertical resolution. 
To address this, 3D-ICE employs a layer subdivision algorithm that profiles and analyzes the vertical thermal resistance distribution, and subdivides layers that contributes a disproportionately large resistance into multiple thinner layers. This strategy can decrease the resistance for a single layer, guide the resistance distribution closer to a uniform distribution, and better capture temperature gradients in the vertical direction. 

Following layer division, mesh generation is performed. Uniformly meshing the entire chip at fine granularity would explode the problem size, especially for stacks with many layers. To have a better trade-off between accuracy and efficiency, 3D-ICE utilizes a local refinement strategy. It first generates a coarse mesh and runs an initial simulation with power maps imported from power models, then identifies regions where hotspots or large temperature gradients occur. The mesh in these regions is adaptively refined to improve the temperature fidelity without excessively increasing the problem size. Furthermore, parallel acceleration on CPUs is employed to speed up the construction of the RC network and the solution of equation. 

Future improvements in the aforementioned context involve expanding the geometry import interface to support more industrial file formats for better flexibility. Moreover, predicting potential hotspot regions directly from power maps, rather than relying on a preliminary simulation, could streamline the mesh refinement process. Additionally, leveraging more efficient hardware acceleration, such as GPU acceleration, could further boost the speed and scalability of thermal analysis for large-scale structures.  

\subsubsection{Physics-informed Neural networks (PINNs)}
In addition to compact thermal models, various neural networks have also been explored for thermal modeling. However, traditional purely data-driven neural network models often require extensive training data and may struggle to enforce physical laws or boundary conditions, which limit their accuracy and generalizability. PINNs are promising for addressing these challenges by incorporating the governing equation and boundary conditions into the learning process. This integration enables unsupervised training without a large ground-truth dataset, and yields solutions that inherently satisfy key physics constraints. Recent studies have begun applying PINNs to chip thermal analysis. For example, ThermPINN~\cite{chen2023fast} uses a discrete cosine series expansion within a PINN to enforce boundary conditions and accelerate convergence during training. Another example is DeepOHeat \cite{liu2023deepoheat}, which employs a deep operator network to achieve real-time 3D temperature predictions. By bridging the gap between data-driven models and equation-based solvers, PINNs offer a promising avenue for fast and physics-consistent thermal predictions, especially in scenarios where data is scarce or expensive to obtain. 
\bhl{However, PINNs in this domain still face practical challenges, including the long training time, instability issues, and weak cross-design transferability. Overcoming these hurdles is necessary for PINNs to see a wider adoption in on-chip thermal modeling.}


\subsection{Multi-physics co-simulation}
Accurate evaluation of future MPSoCs will require multi-physics co-simulation frameworks that consider electrical, thermal, and mechanical effects together \cite{ma2024electrical}. Thermal conditions can significantly affect electrical behaviors. For example, higher junction temperatures increase electrical resistivity, leading to a larger IR drop in power delivery networks and a higher current draw. That increased current, in turn, produces more Joule heating, creating a positive feedback loop that accelerates thermal runaway and electromigration \cite{lu2016electrical}. At the same time, spatial temperature gradients induce mechanical stresses as a result of differential thermal expansion. In vertically stacked 3D ICs, layers heating at different rates can expand unevenly, potentially warping the stack and compromising signal integrity or long-term reliability. 

\bhl{
Today's multiphysics simulation toolchains usually couple separate domain-specific solvers in a sequential or iterative manner, which have several limitations: (i) high computational overhead, since each iteration must solve large systems of equations for each physics domain; (ii) numerical stability issues caused by vastly different spatial and temporal scales; and (iii) required field-to-field interpolation if meshes differ between tools. 
Recent research has started to mitigate these issues. For example, adaptive time-stepping algorithms can adjust each domain's simulation time step independently to maintain stability without excessive over-sampling. Another approach is to incorporate reduced-order models or machine learning surrogate models for data exchange between solvers. Despite these advances, developing a robust and scalable multiphysics co-simulation framework for large-scale 2.5D/3D ICs remains an open challenge.
}

\subsection{Power and thermal management for Heterogeneous MPSoCs}
Beyond advancements in modeling and simulation, intelligent runtime management will be critical for future heterogeneous MPSoCs. Modern SoCs increasingly integrate diverse components, such as big and little cores, AI accelerators, and other specialized IPs. These components are tightly coupled through shared memory interfaces, power distribution networks, and heat dissipation pathways, while differing in voltage-frequency scaling profiles, workload and thermal characteristics. Coordinating such diverse subsystems at run time is a complex control problem due to the large state space, encompassing instantaneous workload distribution, temperature conditions, power budgets, and performance targets.

To address this complexity, the most recent work in this area is shifting toward more adaptive, predictive, and machine learning-based management schemes. For example, Huang et al. \cite{huang2024evaluation} propose a multi-agent reinforcement learning (MARL) framework in which each core is controlled by an agent, and the agents collaboratively aim to optimize power efficiency, maintain thermal balance, and meet performance goals. 
\bhl{
In another study, Sikal et al. \cite{sikal2024ml} present a framework that couples DVFS with thermal-aware task migration to relieve hotspots and cache contention in many-core processors with 3D HBM, achieving notable performance gains while respecting temperature limits.
}
Similarly, Maity et al. \cite{maity2025harnessing} develop new lightweight Q-Network controllers for heterogeneous CPU-GPU MPSoCs that select appropriate CPU and GPU voltage-frequency levels and dynamically allocate workloads between the CPU and GPU. 

Overall, future research must continue to extend these previously mentioned approaches into robust and general frameworks that can coordinate the use of multiple control knobs across all components of the next-generations of heterogeneous MPSoCs.  

%% file: bio.tex

\begin{IEEEbiographynophoto}{David Atienza} (M'05-SM'13-F'16) is a professor of Electrical and Computer Engineering, and heads the Embedded Systems Laboratory at EPFL, Switzerland. He earned his Ph.D. in Computer Science and Engineering from UCM, Spain, and IMEC, Belgium. His research focuses on system-level design methodologies for high-end MPSoC servers and data centers, and ultra-low power embedded systems. He has over 450 publications in peer-reviewed journals and conferences, and 14 patents. He serves or has served as Editor-in-Chief (EiC) for IEEE TCAD and ACM CSUR, and as Deputy EiC for IEEE TCASAI. Dr. Atienza is a Fellow of IEEE and ACM.

\begin{IEEEbiographynophoto}{Kai Zhu} received his MSc degree in electrical engineering from Beihang University in 2023. He is currently pursuing the PhD degree with the Embedded Systems Laboratory (ESL), École Polytechnique Fédérale de Lausanne (EPFL). His research interests focus on the thermal modeling and management of MPSoCs.
\end{IEEEbiographynophoto}

\begin{IEEEbiographynophoto}{Darong Huang} received his Ph.D. degree from École Polytechnique Fédérale de Lausanne (EPFL) in 2025. He is currently a postdoctoral researcher at the Embedded Systems Laboratory (ESL), EPFL, where he focuses on thermal and reliability management for Multi-Processor System-on-Chip (MPSoC) platforms, as well as multi-objective optimization and management of cloud servers.
\end{IEEEbiographynophoto}

\begin{IEEEbiographynophoto}{Luis Costero} received the PhD degree in computer sciences from the Complutense University of Madrid (UCM), in 2021.  Since 2022, he is an assistant professor with the Department of Computer Architecture and System Engineering, at Complutense University of Madrid, Spain. His main research areas involve high performance computing, heterogeneous architectures, power consumption and real-time resource management.
\end{IEEEbiographynophoto}

\end{IEEEbiographynophoto}